\documentclass[12pt]{article}
\pdfoutput = 1
\usepackage{amsmath,amssymb,bm,epsfig,afterpage}
\usepackage{cite}
\usepackage{here}
\usepackage{color}
\usepackage{tabularx}

\renewcommand{\thefootnote}{\fnsymbol{footnote}}
\newcommand{\vev}[1]{{\langle{#1}\rangle}}

\newcommand{\bmat}{\begin{pmatrix} } 
\newcommand{\emat}{\end{pmatrix} } 
 
\newcommand{\br}[2]{\text{Br}({#1}\to{#2})}  
\newcommand{\ev}[1]{\langle{#1}\rangle} 
\newcommand{\order}[1]{\mathcal{O}\left(#1\right)}
\newcommand{\qqad}{\quad\quad\ }
\newcommand{\brbsg}{\text{Br}(b \to s \gamma)}
\newcommand{\brbsm}{\text{Br}(B_s \to \mu^+\mu^-)}
\newcommand{\mhu}{m^2_{H_u}}

\allowdisplaybreaks[3]
\newcolumntype{Y}{&gt;{\centering\arraybackslash}X} 

\begin{document}
\title{
\begin{flushright}
\begin{minipage}{0.2\linewidth}
\normalsize
UT-17-32 \\*[50pt]
\end{minipage}
\end{flushright}
{\Large \bf 
Analysis of the TeV-scale mirage mediation \\ with heavy superparticles 
\\*[20pt]}}
\author{
Junichiro~Kawamura$^a$\footnote{
E-mail address: kawamura@hep-th.phys.s.u-tokyo.ac.jp} \ and
Yuji~Omura$^b$\footnote{
E-mail address: yujiomur@eken.phys.nagoya-u.ac.jp}\\*[20pt]
{\it \normalsize 
$^a$Department of Physics, University of Tokyo, 
Tokyo 113-0033, Japan} \\
{\it \normalsize 
$^b$Kobayashi-Maskawa Institute for the Origin of Particles and the Universe (KMI), } \\
{\it \normalsize
Nagoya University, Nagoya 464-8602, Japan} \\*[50pt]
}
\date{
\centerline{\small \bf Abstract}
\begin{minipage}{0.9\linewidth}
\medskip 
\medskip 
\small
We discuss effective models derived from a supersymmetric model whose
mediation mechanism of supersymmetry (SUSY) breaking is namely mirage mediation.
In this model, light higgsino mass, that is required by the natural realization of the electroweak scale, 
is achieved by the unification of the soft SUSY breaking parameters at the low scale. 
Besides, we find that extra Higgs fields are also possibly light in some cases.
Then, the effective model is a two Higgs doublet model (2HDM) with higgsinos,
and it is distinguishable with namely type-II 2HDM which is widely discussed. 
In this paper, we study the mass spectrum of SUSY particles and the extra Higgs fields,
and summarize the phenomenology in the effective model.
We survey the current experimental bounds from the LHC and the dark matter
experiments as well as the flavor physics. 
Then, we point out the expected mass scale of the SUSY particles
and reveal the future prospects for the direct and indirect searches.
We also discuss the difference between our effective model and
the 2HDM in the bottom-up approach.
\end{minipage}
}

\begin{titlepage}
\maketitle
\thispagestyle{empty}
\clearpage
\tableofcontents
\thispagestyle{empty}
\end{titlepage}

\renewcommand{\thefootnote}{\arabic{footnote}}
\setcounter{footnote}{0}

\section{Introduction}
One elegant explanation for the origin of the electroweak (EW) scale is given by the supersymmetric extension of the Standard Model (SM)~\footnote{
See for a reviews, e.g.~\cite{Martin:1997ns,Chung:2003fi} 
}.
In the Minimal Supersymmetric Standard Model (MSSM), the superpartners
of the SM particles are introduced and quadratic divergence is canceled out in the
Higgs mass squared. 
The natural realization of the EW scale predicts the superpartners to be the EW-scale, 
so that a lot of efforts have been devoted to discover the new particles.
The latest LHC results on the supersymmetry (SUSY) search, however, 
show that SUSY particles do not exist below a few TeV in a simple scenario. 
In addition, the Higgs discovery around the 125 GeV mass~\cite{Aad:2015zhl} 
indirectly constraints the SUSY scale, since the MSSM
predicts that the mass of a neutral Higgs particle is much lower than 125 GeV without large radiative corrections.
Then, we might conclude that the SUSY scale is much higher than the EW scale.

On the other hand, it is true that the relation between the SUSY scale and the realization of the EW scale is not so simple in the MSSM. In fact, we can find some explicit SUSY models that are able to be consistent with 
the experimental results and explain the origin of the EW scale naturally.
One simple way is to consider the MSSM with non-universal gaugino masses 
at the unification scale ($\sim 10^{16}$ GeV)~\cite{Abe:2007kf,Abe:2012xm}. 
A specific mass ratio of wino to gluino can realize the EW scale naturally,
even if gluino is heavy. 
Then, we can evade the strong bounds from the direct SUSY search
and explain the 125 GeV Higgs mass~\cite{Abe:2015xva,Kawamura:2016drh}.  
The mass ratio is non-trivial, 
but it is known that such a unique mass spectrum is predicted by a mediation mechanism:
namely mirage mediation~\cite{Choi:2006xb,Kitano:2005wc}. 
In this mechanism, 
the moduli mediation~\cite{Kobayashi:1994eh,Brignole:1995fb,Brignole:1997dp} 
and the anomaly mediation~\cite{Randall:1998uk,Giudice:1998xp} are compatible,  
and the renormalization-group (RG) correction of the moduli-mediation contribution 
is canceled by the anomaly-mediation.
Phenomenology of the mirage mediation have been studied 
before the Higgs boson discovery~\cite{Choi:2006im,Cho:2007fg,Nagai:2007ud} 
and after the discovery~\cite{
Asano:2012sv,Kobayashi:2012ee,Abe:2014kla,Hagimoto:2015tua,Baer:2016hfa}. 
The mirage mediation in the Next-to-MSSM~\footnote{
See for a review, e.g.~\cite{Ellwanger:2009dp}. 
}
has been also studied in Refs.~\cite{Asano:2012sv,Kobayashi:2012ee,Hagimoto:2015tua}.  

In the last few years, 
the LHC run-II excludes light SUSY particles (sparticles);
especially, colored sparticles, namely squarks and gluinos, 
have to be heavier than 2 TeV. 
Therefore it is worth to study the scenarios that  
heavy colored sparticles can be compatible with the natural explanation of the EW scale. 

In this paper, we reconsider the mirage mediation and discuss the phenomenology based on the 
latest experimental results. 
Even if the sparticles, except for the superpartner of the Higgs field (higgsino), are much heavier than the LHC reach, 
the extra Higgs bosons and the higgsino become lighter than TeV scale in our scenario. 
The light particles can be tested by the direct LHC search, flavor experiments, precision measurements of the Higgs boson couplings and dark matter (DM) searches. 
Based on the integrated research,
we point out the expected mass scale of the SUSY particles
and reveal the future prospects for the direct and indirect searches.
We also discuss the difference between our effective model and
the 2HDM in the bottom-up approach.

This paper is organized as follows.
The mirage mediation is briefly reviewed in Section 2, 
and phenomenology of the mirage mediation is discussed in Section 3. 
The results of numerical analysis are shown in Section 4. 
Section 5 is devoted to conclusion.

\section{Mirage mediation}
The mirage mediation is a mixture of the modulus and anomaly mediations. 
Once the mediation mechanism of the supersymmetry breaking 
is specified, the soft SUSY breaking terms, given by
\begin{align}
 -\mathcal{L}_{\rm soft} = \frac{1}{2}M_a \lambda_a \lambda_a 
                                       +\frac{1}{2} m_i^2 |\phi^i|^2 
                                       + \frac{1}{6}A_{ijk}y_{ijk} \phi^i\phi^j\phi^k + \text{h.c. }, 
\end{align}
are determined. 
$\lambda_a$ and $\phi^i$  are gauginos 
in vector supermultiplets and scalar fields in chiral supermultiplets. 
In our notation, we factorize scalar trilinear couplings as $A_{ijk}y_{ijk}$, 
where $y_{ijk}$ is a Yukawa coupling. 
The indices $a$ and $i$ run over the MSSM gauge groups 
and the scalar fields, respectively.

In the mirage mediation, 
the soft parameters at the unification scale $M_U$ are given by 
\begin{align} 
 M_a    (M_U)=& M_0\left[1
                                  +\frac{b_a}{16\pi^2}g_0^2\alpha\ln{\frac{M_p}{m_{3/2}}}\right], \\
 A_{ijk}(M_U) =& - M_0\sum_{l=i,j,k}\left[c_l-\frac{\gamma_l}{16\pi^2}
                                                           \alpha\ln{\frac{M_p}{m_{3/2}}} \right], \\
\label{eq-softmass}
 m_i^2  (M_U) =&M_0^2 \left[
            c_i-\frac{1}{32\pi^2} \frac{d\gamma_i}{d\ln Q} 
            \left(\alpha\ln{\frac{M_p}{m_{3/2}}}\right)^2\right. \\ \notag 
           &+\left.\frac{1}{4\pi^2}
            \left(\frac{1}{4}\sum_{j,k}|y_{ijk}|^2 \sum_{l=i,j,k}c_l 
            -\sum_a g_a^2 C_2^a(\phi^i) \right)\alpha\ln{\frac{M_p}{m_{3/2}}}\right],    
\end{align} 
The first and second terms in these expressions correspond 
to the modulus and the anomaly mediation, respectively. 
The last term in Eq.~(\ref{eq-softmass}) comes from 
both of the mediation mechanisms. 
We parametrize the overall size of modulus mediation by $M_0$ 
and each $c_i$ describes the ratio of the scalar mass parameter of $\phi_i$
to $M_0$. 
The size of $c_i$ depends on the coupling of $\phi_i$ with the modulus.
Note that $c_i$ is discrete in the string models.  
$g_0$ is the unified gauge coupling at $M_U$ and  
$C_2^a(\phi^i)$ is the quadratic Casimir for a scalar $\phi^i$. 
$b_a = (33/5, 1, -3)$ is the beta-function coefficients for the MSSM gauge couplings. 
$\gamma_i$ is an anomalous dimension for a scalar $\phi^i$ and given by 
\begin{align}
 \gamma_i = 2\sum_a g_a^2 C_2^a(\phi^i) -\frac{1}{2} \sum_{j,k} |y_{ijk}|^2.  
\end{align}
$d\gamma_i/d\ln{Q}$ is a derivative of $\gamma_i$ 
with respect to the logarithmic of the renormalization scale $Q$.

We parametrize the ratio of the anomaly mediation to the modulus mediation 
by $\alpha$ and it is defined as 
\begin{align}
 \alpha \equiv \frac{m_{3/2}}{M_0 \ln{(M_p/m_{3/2})}}, 
\end{align}
where $M_p \simeq 2.4 \times 10^{18}$ GeV is the Planck mass 
and $m_{3/2}$ is the gravitino mass. 
This parametrization is motivated 
by the KKLT-type moduli stabilization 
scenario~\cite{Choi:2004sx,Choi:2005ge,Choi:2005uz}.   
Note that $\ln{(M_p/m_{3/2})} \sim \order{8\pi^2}$ 
and compensate the loop suppression factors 
appeared in the anomaly mediated contributions. 
$\alpha = 1$ is realized in the original KKLT-setup~\cite{Kachru:2003aw}, 
but various rational values can be obtained 
in the similar setups~\cite{Choi:2006xb,Abe:2005rx,Abe:2006xp}. 
An important fact is that 
$\alpha$ could be determined by rational parameters, 
such as the winding number of D-branes and
the number of fluxes which generates moduli potential. 

A remarkable feature of the mirage mediation
is the unification of some parameters at the low scale ({\it mirage unification})~\cite{Choi:2006xb,Kitano:2005wc}.  
Assuming that the moduli-mediation contributions satisfy
\begin{eqnarray}
\label{eq-mircond}
 c_i+c_j+c_k = 1, 
\end{eqnarray}
the anomaly-mediation contributions cancel out the RG corrections of the modulus-mediation contributions
at a scale ($M_{\rm mir}$). 
The unification scale, namely {\it mirage scale}, is given by 
\begin{eqnarray}
\label{eq-Mmir}
M_{\rm mir} = \frac{M_U}{(M_{p}/m_{3/2})^{\alpha/2}},    
\end{eqnarray}
and the soft parameters at $M_{\rm mir}$ are estimated as
\begin{eqnarray}
 M_a(M_{\rm mir})    = M_0,\ 
 A_{ijk}(M_{\rm mir})= -M_0,\ 
 m_i^2  (M_{\rm mir})= c_i M_0^2. 
\label{eq-softmir}
\end{eqnarray}
If $\alpha\sim 2$, the mirage scale is around TeV-scale. 
In our analysis, 
we assume $M_{\rm mir} = M_{\rm SUSY} \equiv \sqrt{m_{Q_3}m_{u_3}}$, 
where $m_{Q_3}, m_{u_3}$ are soft masses for left-handed and right-handed top 
squarks, respectively.

Equation~(\ref{eq-softmir}) shows that $m_{H_u}^2$ at the {\it mirage scale} 
vanishes at the leading order if $c_{H_u} = 0$ is satisfied.   
This means that $|\mu|^2 \simeq m_{H_u}^2$ is EW-scale 
even when the other scalar masses, which are estimated as $\mathcal{O}(M_0)$,  
are much larger than the EW scale. 
If the {\it mirage condition} is satisfied 
and the A-term is not larger than the scalar masses for the top squarks,  
$M_0$ need to be larger than about 7 TeV 
in order to explain the SM-like Higgs boson mass as discussed later.   
Thus, the 125 GeV Higgs boson mass and heavy sparticles are achieved, 
while the EW scale is realized without fine-tunings 
in the TeV-scale mirage mediation.

In our analysis, we assume
\begin{eqnarray}
 \label{eq-choiceC}
 c_Q=1/2,\ c_{H_u} = c_{H_d}= 0, 
\end{eqnarray} 
where $c_Q$ is for all scalar particles other than $H_{u,d}$.
Note that $c_{H_d} = 0$ has to be satisfied 
in order to realize the {\it mirage unification} 
when $\tan\beta$ is large
and the bottom and tau Yukawa couplings are also sizable 
adding to the top Yukawa coupling.

As pointed out in Ref.~\cite{Choi:2006xb}, 
the sub-leading corrections to the mass squared become important
for parameters that vanish at the tree-level. 
Hence, we assume that $ m_{H_{u,d}}^2$ is small but not vanishing at the mirage scale:
\begin{eqnarray}
  m_{H_{u,d}}^2(M_{\rm mir})=\delta m^2_{H_{u,d}}
                                              =\mathcal{O}\left(\frac{M^2_0}{8\pi^2}\right). 
\end{eqnarray} 
The sub-leading correction would come from 
the fluctuation of the mirage scale from $M_{\rm SUSY}$,   
higher-loop corrections in both the MSSM and UV-models,  
and sub-leading corrections in moduli stabilizations. 
Those corrections are expected to be small, so that  
only the higgsino and the extra Higgs bosons
are below sub-TeV, while all the other sparticles are expected to be heavier than the sub-TeV scale.

\section{Phenomenology}
In this section, we study phenomenology when the soft parameters are given by 
\begin{align}
 M_a(M_{\rm SUSY}) = - A_{ijk}(M_{\rm SUSY}) = M_0,\ 
 m_{i}^2(M_{\rm SUSY})= \frac{M_0^2}{2}, 
\end{align}
and 
\begin{align}
 m_{H_{u,d}}^2 (M_\text{SUSY})= \mathcal{O}\left(\frac{M_0^2}{8\pi^2}\right).  
\end{align}
This alignment is predicted by the TeV-scale mirage mediation, as discussed in previous section. 
We have the following parameters:  
\begin{align}
 \tan\beta,\ M_0,\ m_A,\ \mu.  
\end{align} 
Note that the CP-odd Higgs boson mass $m_A$ and the $\mu$-parameter 
are treated as input parameters instead of 
soft parameters $m^2_{H_{u}}$ and $m^2_{H_d}$. 
The size of modulus mediation $M_0$ is fixed to explain the SM-like Higgs boson mass.

\begin{table}[hp!]
 \centering 
 \caption{Values of parameters, masses, widths, branching ratios, 
production cross section of $H/A$ in association with a b-quark, 
$\kappa_{b,\tau}$, flavor observables and DM observables 
at benchmark points (a)-(d). } 
 \vspace{0.5cm} 
 \begin{tabular}{c|cccc}\hline 
  parameters\rule[0mm]{0mm}{5mm}&\qqad(a)\qqad\ &\qqad(b)\qqad\ &\qqad(c)\qqad\ 
                                                         &\qqad(d)\qqad\   \\ \hline\hline
  $\tan\beta $                        & 15          &  45        &   30         & 10  \\
  $\mu$ [GeV]                       & -1090     &-1068     &  1000       & 100 \\ 
  $M_0$ [GeV]                       & 9896     & 6783      & 7905      & 26700 \\ \hline
  degree of tuning                  &&&& \rule[0mm]{0mm}{4.5mm} \\ \hline\hline 
  $\Delta_{\mu} $                  & 285.8    & 274.3     &  240.5      & 2.405   \\
  $\Delta_{M_0}$                   & 524.6    & 255.0    &  341.1      & 3489   \\ \hline 
  mass [GeV]  && & &\rule[0mm]{0mm}{4.5mm}\\ \hline\hline 
  $m_h$                                 & 125.09  & 125.09   &125.09    & 125.10 \\
  $m_H$                                & 1477      & 1364     & 1429      & 1490     \\
  $m_A$                                & 1477       & 1364     & 1429      & 1489    \\
  $m_{H^\pm}$                    & 1485      &  1362     & 1431      &  1571    \\
  $m_{\tilde{\chi}^0_1}$      & 1117  & 1088 & 1022 &  105.4 \\
  $m_{\tilde{t}_1}$                & 7004       & 4767     & 5578     &  18975  \\
  $m_{\tilde{t}_2}$                & 7264       & 4980     & 5811     &  19465 \\ \hline 
  Long-Lived Particle search &&&& \\ \hline\hline 
  $\Delta m_+$   [GeV]          & 0.827 &  1.071 & 0.918            & 0.395      \\
  $c\tau$               [mm]          & 0.241  & 0.0806     & 0.156  & 4.20 \\ \hline
  branch/cross section           & &&\rule[0mm]{0mm}{4.5mm} \\ \hline\hline  
  $\br{H}{bb}$                       & 0.801        & 0.872   & 0.854     &  0.609   \\
  $\br{H}{\tau\tau}$               & 0.125        & 0.126  & 0.141      &  0.0971\\
  $\br{H^\pm}{tb}$                & 0.866       & 0.856   &0.860      &  0.886  \\
  $\br{H^\pm}{\tau \nu}$       & 0.131        & 0.142  & 0.138      &  0.101\\
$\sigma(p p \to H/A+b)$ [fb]& 1.837      & 30.4 & 9.19 &  0.706 \\ \hline 
   indirect observables            & &&& \rule[0mm]{0mm}{4.5mm} \\ \hline\hline  
   $\kappa_b $                       &1.011       & 1.014  & 1.012& 1.011  \\ 
   $\kappa_\tau$                    &1.011       & 1.014 & 1.012 &  1.011 \\
   $\br{b}{s\gamma}                %
   \times10^{4}$                      &3.39         &3.41   & 3.40    & 3.38         \\ 
   $\br{B_s}{\mu^+\mu^-}        %
   \times10^9$                         &2.97        &1.92   & 3.08  &   3.02         \\ \hline 
   DM observables                  &&&& \rule[0mm]{0mm}{4.5mm} \\ \hline\hline   
   $\Omega_{\rm thermal}h^2$  &0.119      &0.119 & 0.102&  0.00116   \\ 
   $\ev{\sigma v}_{v=0}           %
\times10^{25} [\text{cm}^3/\text{s}]$&0.0873      &0.0946& 0.105&   3.95          \\ 
   $\sigma_{\rm SI  }                %
     \times10^{11}$ [pb]           &0.614      &1.41    & 2.11&    0.152     \\ 
   $\sigma_{\rm SD}                 %
     \times10^8$     [pb]            &0.769     &1.96    & 1.52&     8.97       \\ 
\hline  
 \end{tabular}
\label{tab-vars}
\end{table}

In this setup, 
all of sparticles, except for higgsinos, reside far above the LHC reach, 
while the Higgs bosons are expected to be around TeV-scale. 
One promising way is the direct search for extra Higgs bosons 
and the higgsino at collider experiments.  
Besides, the extra Higgs bosons lead deviations from the SM predictions 
in some processes such as, $B_s \to \mu^+\mu^-$, $B\to X_s\gamma$ 
and the decay of the 125-GeV Higgs boson. 
The higgsino becomes the lightest supersymmetric particle (LSP) in this setup. 
Hence the neutral component of the higgsino is a good DM candidate 
and the direct and indirect DM searches give significant bounds on our model.

The mass spectrum and the values of observables 
at several benchmark points are summarized in Table~\ref{tab-vars}. 
We calculated the mass spectrum of sparticles and their decays 
by using SuSpect-v2.41~\cite{Djouadi:2002ze} 
and SDECAY~\cite{Muhlleitner:2003vg}  
interfaced by SUSY-HIT-v1.5a~\cite{Djouadi:2006bz}.  
The Higgs boson masses, decays and couplings to the SM fermions 
are calculated by using FeynHiggs-v2.12.2~
\cite{Heinemeyer:1998yj,Heinemeyer:1998np,Degrassi:2002fi,Frank:2006yh,Hahn:2013ria,Bahl:2016brp}.  
$M_0$ should be larger than about 6.8 TeV in order to explain the 
125 GeV Higgs boson mass,  
so that superpartners except higgsino are much higher than the LHC reach.

\subsection{Electroweak symmetry breaking} 
First of all, we analyze the condition for the EW symmetry breaking.
In the MSSM, the Z-boson mass $m_Z$ is related to the parameters of the MSSM as 
\begin{align}
 \frac{m_Z^2}{2} = \frac{\tilde{m}_{H_d}^2-\tilde{m}_{H_u}^2 \tan^2\beta}
                                     {\tan\beta^2-1} -|\mu|^2 
                           \simeq - \tilde{m}_{H_u}^2 - |\mu|^2 , 
\label{eq-EWSB}
\end{align} 
where $\tan\beta \gg 1$ is assumed in the last equality.  
$\tilde{m}^2_{H_{u,d}}$ include corrections from the effective potential, 
\begin{align}
 \tilde{m}^2_{H_{u,d}} = m^2_{H_{u,d}} + \frac{1}{2 \vev{H_{u,d}}} 
                                     \left. \frac{\partial \Delta V}{\partial H_{u,d}} 
                                     \right|_{H_{u,d} \to \vev{H_{u,d}} } 
                                    \equiv  m^2_{H_{u,d}} + \Delta m^2_{H_{u,d}},  
\end{align}
where $\Delta V$ is loop corrections to the effective potential. 

According to the assignment of the modular weights in Eq.~(\ref{eq-choiceC}), 
the $m_{H_u}^2$ at a scale $Q$ can be expressed as~\cite{Choi:2005uz}
\begin{align}
\label{eq-mhusq}
 m_{H_u}^2(Q) = \frac{M_0^2}{4\pi^2} 
                        \left\{ \gamma_{H_u}(Q) -\frac{1}{2} \frac{d\gamma_{H_u}(Q)}{d\ln{Q}} 
                            \ln{\left(\frac{M_\text{mir}}{Q}\right) } \right\} 
                            \ln{\left(\frac{M_\text{mir}}{Q}\right) } 
                            + \delta m^2_{H_u}(Q), 
\end{align}
where the anomalous dimension for the up-type Higgs boson is given by 
\begin{align}
 \gamma_{H_u} = \frac{3}{2}g_2^2+\frac{3}{10}g_1^2-3y_t^2. 
\end{align}
Note that the modulus contribution to the Higgs bosons is absent at $Q=M_{\rm mir}$ 
because of $c_{H_u} = c_{H_d} = 0$. 
The RG effects through the $U(1)_Y$ gauge coupling 
are also vanishing in our setup of the modulus mediation.

Let us discuss the sensitivity of the Z-boson mass with respect to 
the parameters in the mirage mediation~\cite{Barbieri:1987fn}: 
\begin{align}
 \Delta = \max_{a} \Delta_a \equiv \max_{a} \left|\frac{d\ln{m_Z^2}}{d\ln{a}} \right|, 
   \       a = |\mu|^2, M_0^2, \delta m_{H_u}^2.  
\end{align}
Here we do not consider tuning of $\delta m_{H_d}$ and the b-term 
because they are suppressed by $\tan^2\beta$ 
and expected to be small~\cite{Choi:2006xb}.     
The degrees of tuning of the $\mu$-parameter 
and $\delta m_{H_u}^2$ are given by
\begin{align}
 \Delta_\mu \equiv \left|\frac{d\ln{m_Z^2}}{d\ln{|\mu|^2}}\right| 
                   \simeq 2 \frac{|\mu|^2}{m_Z^2},\  
 \Delta_{\delta m^2_{H_u}} \equiv \left|\frac{d\ln{m_Z^2}}{d\ln{\delta m^2_{H_u}}}\right| 
                                            \simeq 2\left|\frac{\delta m_{H_u}^2}{m_Z^2}\right|,    
\end{align}
where we treat $\delta m_{H_u}^2$ as a fundamental parameter 
that is independent of other parameters. 
We also assume that the $\mu$-parameter is a fundamental parameter 
since it is an unique supersymmetry preserving parameter.  
Then, $\Delta_\mu$ increases as $\mu$-parameter increases quadratically. 
We see that $\Delta_\mu \sim \Delta_{\delta m^2_{H_u}}$ 
since the minimization condition Eq.~(\ref{eq-EWSB}) requires 
$|\mu|^2 \simeq \delta m^2_{H_u}$. 

We assume that some suitable mechanism fixes $\alpha$ 
to satisfy $M_{\rm mir} = M_{\rm SUSY}$.  
If we consider the KKLT-like setups behind the mirage mediation, 
the value of $\alpha$ may be fixed by rational parameters 
and it is irrelevant to this tuning argument. 
Thus the size of the modulus mediation $M_0$ is important for the argument.  

In Eq.~(\ref{eq-mhusq}), 
$M_{\rm mir}$ implicitly depends on $M_0$ through Eq.~(\ref{eq-Mmir}). 
When we assume that the K$\ddot{\rm a}$hler potential of the moduli is 
\begin{align}
 K_{\rm moduli} = -3 \ln{(T+\overline{T})},   
\end{align}
and only the modulus $T$ gets non-zero F-term vacuum expectation value, 
the size of the anomaly mediation $F^C/C$ relates to $M_0$ and $\alpha$ as 
\begin{align}
 \frac{F^C}{C} &\equiv \alpha M_0 \ln{ \left(\frac{M_p}{m_{3/2}} \right)}  \\ 
                       & = m_{3/2} + \frac{1}{3} K_T F^T= m_{3/2} - M_0,    
\end{align}
where $ K_T \equiv \partial K/\partial T$.  
%
%
Using above relations, we obtain 
\begin{align}
 \frac{d M_{\rm mir}}{d M_0} = \frac{\alpha}{2}\frac{m_{3/2}}{m_{3/2}+\alpha M_0} 
                                                 \frac{M_{\rm mir}}{M_0} 
                                             \equiv c \frac{M_{\rm mir}}{M_0}. 
\end{align}
Note that $c \sim 1$ 
because $m_{3/2} \gg M_0$ 
and $\alpha \sim 2$ when $M_{\rm mir} \sim M_0$. 

Therefore the sensitivity to $M_0$ is 
\begin{align}
\left| \frac{d\ln m_Z^2}{d\ln M_0^2} \right| 
                              \simeq \frac{M_0}{m_Z^2} \left|\frac{d \mhu}{d{M_0}} \right|
                              \simeq \frac{M_0}{m_Z^2}  \left|
                                  c  \frac{\gamma_{H_u}}{4\pi^2} M_0 
                                  + \frac{d \Delta\mhu}{d{M_0}} 
                                  \right|,  
\end{align}
where we choose $Q=  M_{\rm SUSY} = M_{\rm mir} = M_0/\sqrt{2}$.  
We see that the sensitivity is suppressed by the loop factor. 
In our numerical analysis, 
we evaluate the correction from the 1-loop 
Coleman-Weinberg potential~\cite{Coleman:1973jx,Weinberg:1973ua}  
induced by the third-generation quarks and squarks. 

\begin{figure}[t]
 \centering 
 \includegraphics[width=0.475\linewidth]{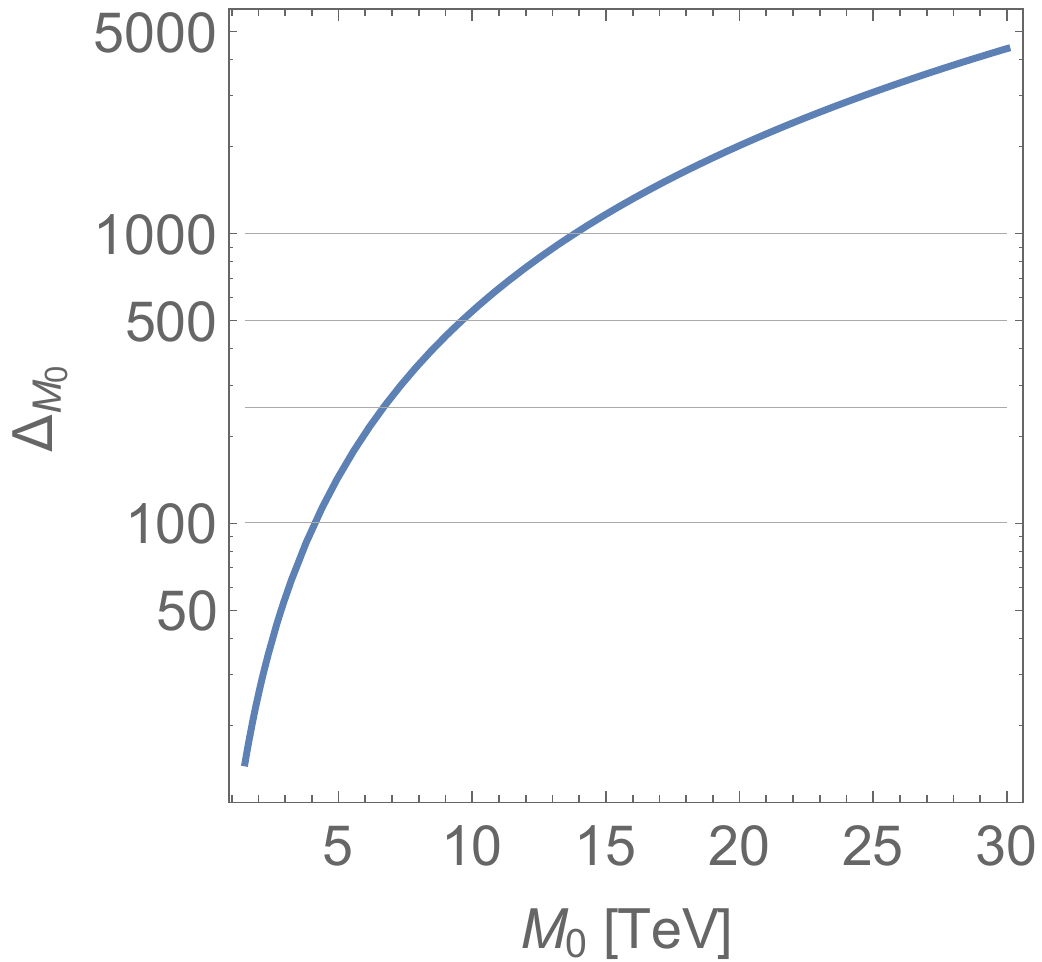}
 \hspace{0.15cm}
 \includegraphics[width=0.48\linewidth]{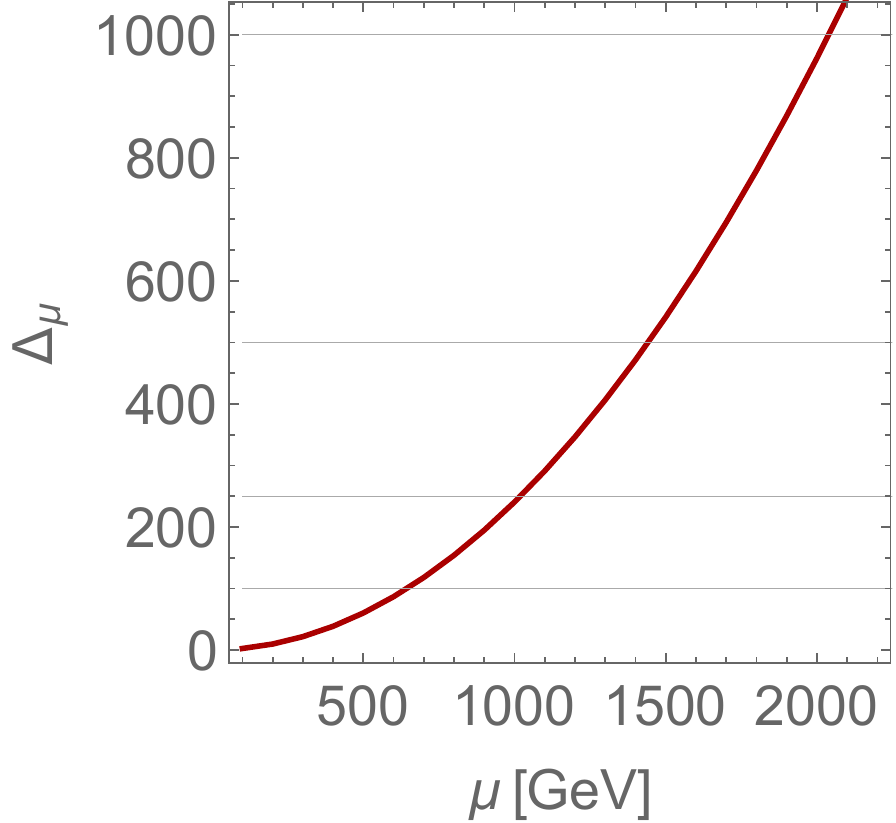}
 \caption{
Degrees of tuning of $M_0$ (left) and $\mu$ (right). 
The gray lines correspond to $\Delta_{M_0,\mu} = 100, 250, 500, 1000$ 
from bottom to top. 
} 
\label{fig-delta}
\end{figure}

Figure~\ref{fig-delta} shows the degree of tuning of $M_0$ and the $\mu$-parameter. 
We see that $\Delta_{M_0} \lesssim 100, 250$ and 1000 
when $M_0 \lesssim 4.6, 6.8$ and 14~TeV, respectively. 
$\Delta_\mu \le 100$  when $|\mu| \lesssim 650$ GeV. 
In the case $\mu \simeq 1.1$ TeV 
where the  thermal relic density of the higgsino saturates the observed value of DM, 
$\Delta_\mu \simeq 290$ and the degree of tuning is about $0.34 \%$. 
Since the 125~GeV Higgs boson mass requires $M_0 \gtrsim 6.0$ TeV 
as we will see later,  
the tuning of $M_0$ becomes severer than that of the $\mu$-parameter 
as long as $|\mu| \lesssim 1.1$ TeV.  
In these regions, 
the electroweak symmetry breaking occurs 
with the parameter tuning about 0.3\%-level. 
This is much better than other scenarios with sparticles heavier than about  5~TeV.

\subsection{Direct LHC search} 
In this setup, 
the extra Higgs bosons are expected to be lighter than the TeV scale 
and they could be discovered by the LHC experiments.  
The neutral Higgs bosons $H/A$ dominantly decay to a pair of bottom quarks 
with a branching fraction $\sim 80\%$
and sub-dominantly decay to a pair of tau leptons 
with a branching fraction $\sim 10\%$. 
When $\tan\beta$ is small and the bottom and the tau Yukawa couplings 
are suppressed,  
the Higgs bosons also decay to a pair of top quarks. 
The searches for $H, A \to \tau\tau$ however give the most stringent bound 
on their masses, since there are large amounts of backgrounds 
for the Higgs bosons decaying to quarks.


The experimental analyses constrain $\sigma(p p \to H/A) \times {\rm Br}(H/A \to \tau\tau)$, 
where the Higgs bosons are produced by the gluon fusion process 
or b-associated process~\cite{ATLAS:Higgs}.  
In our analysis, 
we calculate the production cross section $\sigma(p p \to H/A)$ 
associated with bottom quark  
by using MadGraph-v2.5.4 with 5-flavor scheme~\cite{Alwall:2014hca}.  
We calculate the experimental limits based on the result of Ref.~\cite{ATLAS:Higgs}    
obtained by 36.1 ${\rm fb}^{-1}$ data with the center of energy 
$\sqrt{s}=$13 TeV.

The higgsino can be also detected by the collider experiments. 
There are two neutral 
and two charged components in the higgsino. 
Their masses tend to be mostly degenerate and 
mass differences are smaller than sub-GeV in our scenario. 
The mass differences come from the mixing with gauginos, 
and decrease as gaugino masses increase. 
The mass difference 
$\Delta m_+ \equiv m_{\tilde{\chi}_1^\pm}-m_{\tilde{\chi}_1^0}$  
is smaller than about $\order{0.1\rm GeV}$ for $M_0 \gtrsim 6.0$ TeV at the tree-level. 
$\Delta m_+$ get smaller for larger $M_0$, 
but there are radiative corrections 
from SM gauge boson loops~\cite{Thomas:1998wy,Cirelli:2005uq}. 
The radiative correction for the mass difference is given by 
\begin{align}
\label{eq-delrad}
 \Delta m_\text{rad} = \frac{\alpha_2}{8\pi} m_Z s_W^2 
                                   f\left(\frac{m_Z}{m_{\tilde{\chi}}}\right), 
\end{align} 
where 
\begin{align}
 f(r) = 2r^3\ln{r}-2r+\sqrt{r^2-4}(r^2+2)\ln{\left(\frac{r^2-2-r\sqrt{r^2-4}}{2}\right)}. 
\end{align}
Thus the size of the correction is 
 $\Delta m_\text{rad} \simeq \alpha_2 m_Z s_W^2 /2 \sim  350$ MeV
and would dominate the mass difference. 
The values of $\Delta m_+$ including the loop effect in Eq.(\ref{eq-delrad}) 
are shown in Table~\ref{tab-vars}. 
The LEP experiment gives the most stringent bound 
on such degenerate higgsinos~\cite{Heister:2002mn}: 
the higgsino has to be heavier than about 90 GeV. 

On the other hand,
the expected signals are so weak that these are buried under the backgrounds 
in hadron collider experiments 
and the mass limits could not be severer than the one at the LEP. 
Recently, the higgsino search exploiting disappearing tracks are proposed 
in Refs.~\cite{Mahbubani:2017gjh,Fukuda:2017jmk}. 
When the mass difference 
$\Delta m_+ \equiv m_{\tilde{\chi}_1^\pm}-m_{\tilde{\chi}_1^0}$ 
is smaller than about 0.6 GeV,  
the chargino dominantly decays to $e\nu_e$, $\mu \nu_\mu$ and $\pi^+$ 
and the lightest neutralino. 
The partial decay widths are given by~\cite{Thomas:1998wy,Chen:1996ap} 
\begin{align}
 \Gamma(\tilde{\chi}_1^+ \to \tilde{\chi}^0_1 l \nu_l) 
  &= \frac{2 G_F^2}{15\pi^3}(\Delta m_+)^5 \sqrt{1-x_l^2}\ \mathcal{P}(x_l), 
        \quad (l = e, \mu), \\ 
 \Gamma(\tilde{\chi}_1^+ \to \tilde{\chi}^0_1 \pi^+)  
  &= \frac{2 G_F^2 f_\pi^2 \cos^2\theta_C }{\pi}(\Delta m_+)^3 
        \left(1-\frac{x_\pi^2}{2}  \right) \sqrt{1-x_\pi}, 
\end{align}
where $x_{l,\pi} = m_{l,\pi}/\Delta m_+$, 
$G_F$ is the Fermi constant, $f_\pi \sim 91.9$ MeV is the pion decay constant, 
$\theta_C$ is the Cabbibo angle and 
\begin{align}
 \mathcal{P}(x) = 1-\frac{9}{2} x^2 - 4x^4+\frac{15 x^4}{2\sqrt{1-x^2}} 
                            \tanh^{-1}\sqrt{1-x^2}.  
\end{align}
The decay length of the chargino could be longer than $\mathcal{O}(0.1 {\rm cm})$ 
in our setup,  
so that the future hadron collider experiments would give stronger bounds than 
the LEP experiment. 
The decay lengths calculated from the above three decay modes are shown 
in Table~\ref{tab-vars}. 
The expected exclusion limits at the HL-LHC and the 33-TeV hadron collider 
are shown in the next section,   
referring the result shown in Ref.~\cite{Fukuda:2017jmk}. 
We refer the most optimistic cases in the future collider experiments, 
where there is no background events 
and the location of the second layer of a pixel detector is at a radius 3 cm 
in the 33-TeV collider. 


\subsection{Precision Higgs coupling measurement}
The light extra Higgs bosons change the couplings between
the 125-GeV Higgs boson and SM particles; especially, couplings with  bottom quark and tau lepton can be largely deviated from the SM prediction~\cite{Hall:1993gn,Hempfling:1993kv,Carena:1994bv,Carena:1998gk,Eberl:1999he}.  
There are recent studies about the precision
measurements of the Higgs couplings ~\cite{Endo:2015oia,Bae:2015nva,Kakizaki:2015zva}.  
Now we define a ratio of a coupling in the MSSM to the one in the SM, 
\begin{align}
 \kappa_f \equiv g_{h\bar{f}f}^{\rm MSSM}/g_{h\bar{f}f}^{\rm SM},\ (f = b,\tau), 
\end{align}
where $g_{h\bar{f}f}^{\rm I}$ ($I=$ MSSM, SM) is the coupling between the 125-GeV Higgs ($h$) and the SM fermions ($f$):
\begin{align}
 -\mathcal{L}^I_{h\bar{f}f} = g^I_{h\bar{f}f} h \bar{f}f. 
\end{align}
$\kappa_f$ are written as 
\begin{align}
 \kappa_f = -\left(\frac{\sin\alpha_h}{\cos\beta}\right)
                    \frac{1-\Delta_f\cot\alpha_h \cot\beta}{1+\Delta_f}, 
\end{align}
where $\alpha_h$ is a mixing angle between two CP-even Higgs bosons. 
The first factor comes from purely Higgs boson mixing, 
and then it also exists in the type-II two Higgs doublet model. 
The factor, $\Delta_f$, comes from radiative corrections induced by sparticles. 
$\Delta_b,\ \Delta_\tau$ are approximately given by 
\begin{align}
 \Delta_b &\simeq \left[\frac{2\alpha_s}{3\pi}M_3\mu 
                                  I\left(m^2_{\tilde{b}_1},m^2_{\tilde{b}_2},M_3^2\right) 
                                 +\frac{y_t^2}{16\pi^2}\mu A_t 
                                  I\left(m^2_{\tilde{t}_1},m^2_{\tilde{t}_2},  \mu^2\right) 
                        \right]\tan\beta, \\
\Delta_\tau &\simeq -\frac{3\alpha_2}{8\pi} M_2 \mu 
                               I(m^2_{\tilde{\tau}_L},M_2^2,\mu^2) \times \tan\beta, 
\end{align}
where $ I(a,b,c)$ is defined as
\begin{align}
 I(a,b,c) = - \frac{ab\ln{a/b}+bc\ln{b/c}+ca\ln{c/a}}{(a-b)(b-c)(c-a)}. 
\end{align}
Note that these corrections are enhanced by $\tan\beta$. 
We calculate $\kappa_f$, using FeynHiggs 2.12.2 
which includes the re-summation of the $\Delta_b$ corrections~\cite{Hofer:2009xb} 
and the two-loop SUSY QCD corrections~\cite{Noth:2008tw}. 
We have checked that the other couplings of the SM-like Higgs boson
are very close to the SM value. 
The current accuracy of the Higgs coupling measurements at the LHC is 
$\mathcal{O}(10\%)$~\cite{Aad:2015gba,Khachatryan:2014jba}.
The future sensitivity may reach a few $\%$ in the HL-LHC 
and the measurement may be more accurate than $1 \%$ 
in the future lepton collider experiments 
such as ILC and TLEP~\cite{Peskin:2012we,Dawson:2013bba}.  
The HL-LHC will be able to measure $\kappa_\tau$ more precisely, 
while $\kappa_b$ can be determined more precisely 
at the lepton collider experiments. 


\subsection{Flavor physics}
In our model, the flavor violating couplings involving sparticles are only given by the CKM matrix.
Then, we can evade the strong bounds from flavor physics. 
It is known that the stringent bound on this kind of model comes from the rare $B$ meson decays: $B \to X_s \gamma$ \cite{Hermann:2012fc,Misiak:2015xwa,Misiak:2017bgg}
and $B_s \to \mu^+\mu^-$ \cite{Altmannshofer:2012ks,Altmannshofer:2017wqy}.
If there is only one extra Higgs doublet that couples to the right-handed down-type quarks as in the type-II 2HDM,
there is an one-loop correction involving charged Higgs which does not depend on $\tan \beta$.

In our supersymmetric model, there is also a superpartner of Higgs field, namely higgsino, below 1 TeV.
There may be a cancellation between the charged Higgs loop and the higgsino loop \cite{Chen:2009cw}.


We calculate $\text{Br}(B_s\to X_s\gamma)$ 
and $\text{Br}(B_s\to\mu^+\mu^-)$ 
by using micrOmegas-4.3.2~\cite{Belanger:2013oya,Belanger:2014vza, Belanger:2008sj}. 
We adopt the experimental values as follows:
$\br{b}{s\gamma}_{\rm exp}        = (3.32\pm0.16)\times10^{-4}$ for $E_{\gamma}>1.6$ GeV~\cite{Amhis:2016xyh} and  
$\br{B_s}{\mu^+\mu^-}_{\rm exp} = (3.00\pm0.55)\times10^{-9}$~\cite{CMS:2014xfa,Aaij:2017vad,Altmannshofer:2017wqy}.   
The SM predictions we use are 
$\br{b}{s\gamma}_{\rm SM} = (3.36\pm0.23)\times10^{-4}$ for $E_{\gamma}>1.6$ GeV~\cite{Misiak:2015xwa} and
$\br{B_s}{\mu^+\mu^-}_{\rm SM} = (3.60\pm0.18)\times10^{-9}$~\cite{Altmannshofer:2017wqy}.  

\subsection{DM physics}
The neutral component of the higgsino is the LSP and a good candidate for the dark matter in our scenario~\footnote{
See for a review of the supersymmetric dark matter, e.g.~\cite{Jungman:1995df}. 
}. 
There are several ways to produce the higgsino in the early universe. 

If the mirage mediation is realized by the KKLT-like setup 
and the moduli and the gravitino masses are below PeV-scale,  
late-time decays of these particles produce the higgsino. 
However, 
it is known that the higgsino dark matter produced 
in such a way tends to overclose the universe~\cite{Endo:2006zj}.  
Hence, 
the production of the higgsino should be suppressed,  
or  the produced higgsino should be diluted 
by e.g. thermal inflation~\cite{Lyth:1995hj,Lyth:1995ka}.
If the SUSY breaking scale is enough heavy $M_0 \gtrsim \mathcal{O} (10)$ TeV 
and the higgsino mass is enough light  $m_\chi \lesssim \mathcal{O} (100)$ GeV 
that the higgsinos produced by the late-time decays annihilate enough efficiently, 
the overclose problem could be circumvented 
and the relic density could be explained
by the higgsino produced by the non-thermal way~\cite{Aparicio:2016qqb}.  
Otherwise, the model should be extended to have 
another light supersymmetric dark matter, 
such as axino~\cite{Choi:2009qd,Nakamura:2008ey}.  
In this case, 
the dark matter is no longer the higgsino, 
while the dark matter could be the higgsino dark matter in the other cases.  

It would be possible that 
the relic density of the higgsino is explained 
by the usual thermal freeze-out mechanism, 
when the higgsino mass is about 1.1 TeV~\cite{Cirelli:2005uq,Cirelli:2007xd}. 
This occurs if the gravitino and moduli fields 
decay earlier than the higgsino freezing-out, 
although the decay widths of the gravitino and the moduli should be larger than 
the ones naively expected from the KKLT-like setups. 
Another possibility is that  the mass spectrum of the mirage mediation 
is realized by some other mechanisms than the KKLT-like setup.  
This scenario looks interesting from the bottom-up point of view. 

In Fig. \ref{fig-omega}, 
we consider the case that the higgsino dark matter is only thermally produced and 
saturates the observed relic density : $\Omega_{\rm DM} h^2 = 0.1188\pm0.001$~\cite{Ade:2015xua}. 
In other case, we study the constraints from the direct and indirect detections of the dark matter, assuming that the observed relic density is fully occupied by the higgsino dark matter. 
We calculate the thermal relic density of the higgsino, 
spin-independent cross section and the annihilation cross section 
by using micrOmegas-4.3.2~\cite{Belanger:2013oya,Belanger:2014vza, Belanger:2008sj}.

Note that the discussion about the DM in this Subsection would become irrelevant, if the observed DM relic density is explained by some particle(s) other than the higgsino. 
If the higgsino does not dominate the dark matter density,  
the limits from the DM observations are relaxed by the rescaling factor, $\xi \equiv \Omega_\chi/\Omega_{\rm DM}$,
where $\Omega_\chi$ is the higgsino density and $\Omega_{\rm DM}$ is the total DM density.
The cross section for the direct detection should be rescaled by the factor $\xi$ 
and the cross section for the indirect detection, which observes the cosmic rays originated from the annihilation of the higgsinos, should be rescaled by $\xi^2$, when our predictions are compared with the experimental results.

There are studies about the DM in the mirage mediation and other similar setups~\cite{Choi:2006im,Abe:2007je,Holmes:2009mx}. In the previous works, the size of modulus mediation is below sub-TeV, so that
the 125-GeV Higgs boson mass can not be realized.  
Furthermore, 
the constraints from the direct detections now become very strong, 
so that the gaugino masses should be heavier than sub-TeV 
as far as the higgsino dominates the DM relic density.  
This situation is similar to the one in Ref.~\cite{Kawamura:2017amp}, where
the phenomenology of the higgsino DM is studied in the Non-Universal Gaugino Mass scenario. 
The most important difference with the previous work in Ref.~\cite{Kawamura:2017amp} is that 
the sparticle masses are much heavier in our scenario,  
because the size of A-term is fixed by the mediation mechanism at relatively small values 
and the heavy top squark is required to explain the Higgs boson mass around 125 GeV. 
Moreover, there are relatively light exotic Higgs bosons. 

The direct-detection experiments constrain a cross section of DM-nucleon scattering;
especially, the spin-independent cross section gives 
stringent bounds on the parameter space of the MSSM. 
The spin-independent cross section for the higgsino DM is approximately given by, 
\begin{align}
\label{eq-sigSI}
\sigma^{\rm SI}_{N} \simeq &  
 \frac{g^2}{4\pi} \frac{m_N^4}{m_h^4m_W^2} 
 \left(1+\frac{m_N}{m_\chi} \right)^{-2}  \\ \notag
 &\times \left[\left(\sum_{q=u,
d,s} f_q + \frac{2}{9}f_{TG}\right)\lambda_{h\chi\chi} 
       -\frac{m_h^2}{m_H^2} 
        \left(\sum_{q=u,d,s} f_{T_q} t_{q} +\frac{2}{27}f_{TG} \sum_{q=c,b,t} t_q \right)
        \lambda_{H\chi\chi}  \right]^2  \\
 =& \frac{g^4 }{16\pi} \frac{m_N^4}{m_h^4} 
       \left(1+\frac{m_N}{m_\chi}\right)^{-2}  
       \left( \frac{1}{M_2-|\mu|}+ \frac{t_W^2}{M_1-|\mu|} \right)^2 \\ \notag 
   &\times\left[\left(\sum_{q=u,d,s}f_q+\frac{2}{9}f_{TG}\right)
          (1\pm s_{2\beta})
       \pm \frac{m_h^2}{m_H^2} |c_{2\beta} |
        \left(\sum_{q=u,d,s} f_{T_q} t_{q} +\frac{2}{27}f_{TG} \sum_{q=c,b,t} t_q \right) 
       \right]^2, 
\end{align}
where $m_N$ is the nucleon mass, 
$m_N f_{T_q}^N = \langle N|m_q \bar{q}q|N\rangle$ 
and $f_{TG} = 1 - \sum_{q=u,d,s} f_{T_q}$. 
$t_{q} = -\cot\beta $ for $q = u, c, t$ and 
$t_{q} =  \tan\beta $ for $q = d, s, b$. 
$\pm$ corresponds to the relative sign of the gaugino masses and the $\mu$-parameter.
$\lambda_{h\chi\chi},\ \lambda_{H\chi\chi}$ 
are the higgsino-higgsino-Higgs boson couplings, 
\begin{align}
\label{eq-lamhnn}
 \lambda_{h\chi\chi} &= \frac{g}{2} \left(1 + \text{sgn} (\mu M_{1,2}) s_{2\beta}\right ) 
                           c_Wm_Z      \left( \frac{1}{M_2-|\mu|}+ \frac{t_W^2}{M_1-|\mu|} \right), \\
 \lambda_{H\chi\chi}&= \frac{g}{2}\text{sgn} (\mu M_{1,2})  c_{2\beta}
                           c_Wm_Z \left( \frac{1}{M_2-|\mu|}+\frac{t_W^2}{M_1-|\mu|} \right),
\end{align}
where $c_W, t_W, s_{2\beta}$ and $c_{2\beta}$ are short for  
$\cos\theta_W,\tan\theta_W, \sin2\beta$ and $\cos2\beta$, respectively. 
We replaced $\alpha_h$ to $\beta$ 
by taking the decoupling limit $m_A \gg m_h$ 
and drop all contributions from the sparticles.

The higgsino-higgsino-Higgs boson couplings are originated from 
the higgsino-gaugino-Higgs boson couplings in the gauge-basis. 
This means that the couplings are suppressed by the gaugino masses 
as can be read from Eq.~(\ref{eq-lamhnn}).  
The heavy Higgs boson contribution becomes 
constructive (destructive) when $\text{sgn}(\mu M_{1,2}) = +1(-1)$. 
Besides, the light Higgs boson contribution is proportional to 
$1+\text{sgn}(\mu M_{1,2}) s_{2\beta}$. 
Thus the spin-independent becomes significantly large
when $\text{sgn}(\mu M_{1,2}) = +1$. 

The XENON1T experiment~\cite{Aprile:2017iyp,Aprile:2015uzo} 
and the PandaX-II experiment~\cite{Cui:2017nnn} 
give the most severe bounds on the spin-independent cross section. 
The current limit is about $10^{-10}$ pb for $m_\chi \sim 100$ GeV 
and $10^{-9}$ pb for $m_\chi \sim 1100$ GeV. 
There are future experiments such as LZ experiment~\cite{Akerib:2015cja}, 
XENON-nT experiment and so on.   
These experiments will probe wide parameter space 
as long as the spin-independent cross section times the rescaling factor 
$\xi \sigma_{\rm SI}$ is on the so-called neutrino floor~\cite{Billard:2013qya}. 
If $\xi \sigma_{\rm SI}$ is below the neutrino floor, 
the signals of the dark matter are buried under the neutrino background. 
 
If the dark matter is dominated by the higgsino, 
we will observe cosmic rays originated from the higgsino annihilation. 
In most parameter region, 
a higgsino pair annihilates to a pair of W-bosons and of Z-bosons 
through the t-channel higgsino exchange 
except for the case with $m_A \simeq 2 m_\chi$. 
An important fact is that 
these processes are independent of other sparticles masses, 
because these are mediated by the higgsino itself. 
Thus the indirect detections constrain the higgsino mass itself. 
If $m_A \simeq 2 m_\chi$ is satisfied,  
a higgsino pair annihilate into a pair of bottom quarks 
(sub-dominantly tau leptons) 
through the s-channel CP-odd Higgs boson exchange.  
Although this contribution potentially becomes sizable, 
the higgsino-higgsino-Higgs boson coupling is suppressed
by the gaugino masses as in the spin-independent cross section, 
so that  these processes can not dominate the total annihilation process 
unless the Higgs boson mass is at the resonance region precisely. 
We could not observe such enhancement of the annihilation cross section 
in our numerical analysis in next section. 

The current limit on the annihilation cross section comes from 
the AMS-02 experiment~\cite{Aguilar:2016kjl} which detects anti-protons. 
Analyses in Refs.~\cite{Cuoco:2016eej,Cui:2016ppb} 
show the upper limits on the cross section, 
and the higgsino mass lighter than about 500 GeV 
has been already excluded~\cite{Kawamura:2017amp} 
if the higgsino saturates the dark matter density. 
The future experiments, such as the CTA experiment~\cite{Carr:2015hta}, 
would reach the cross section, $\sim 1.0 \times 10^{-26}\ \text{cm}^3/\text{s}$, 
although there is a large uncertainty due to the unknown profiles of the dark matter.

\section{Numerical Analysis} 
\begin{figure}[t]
 \centering 
 \includegraphics[width=0.48\linewidth]{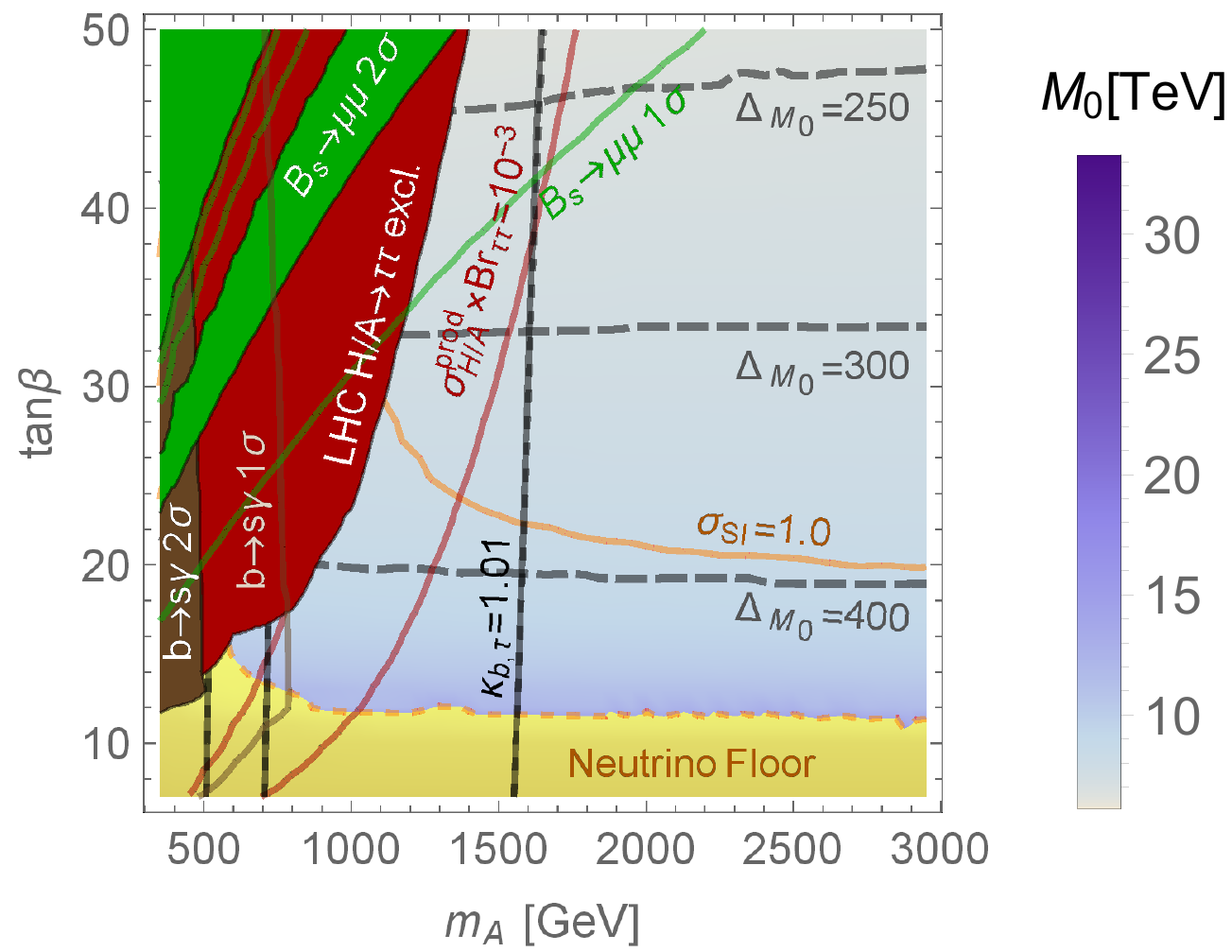}
 \hspace{0.15cm}
 \includegraphics[width=0.48\linewidth]{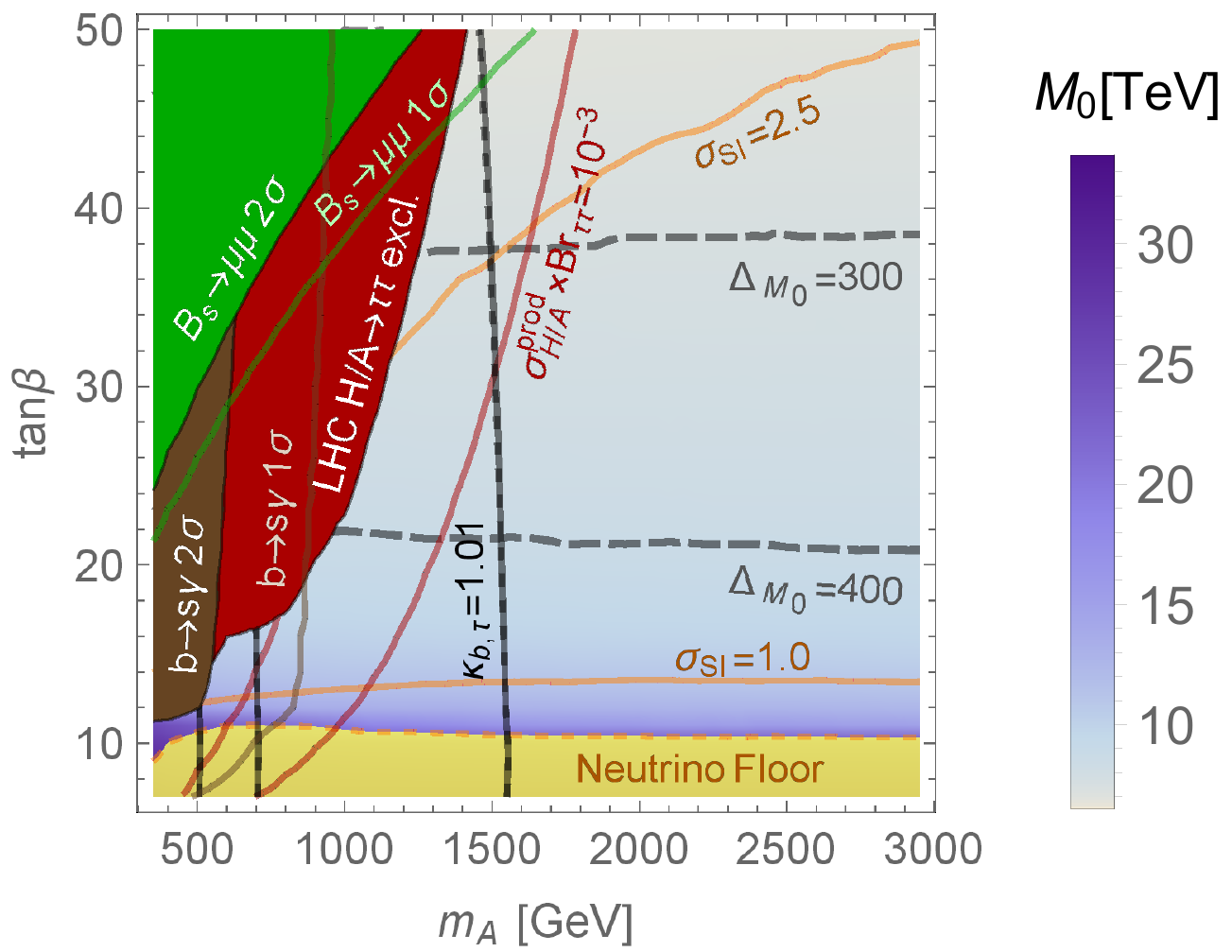}
 \caption{
Values of the observables in the case with $\Omega_\chi h^2$=0.1188. 
$\text{sgn}(\mu M_{1,2}) = -1\ (+1)$ in left (right) panel. 
The color of the background denotes the size of $M_0$. 
The dashed gray lines correspond to $\Delta_{M_0} = 400, 300, 250$ 
from bottom to top. 
The green and brown regions are excluded by $B_s \to \mu^+ \mu^-$ and $b \to s \gamma$ at the 2 $\sigma$ level.
The 1-$\sigma$ exclusion limits are also depicted by the green and brown lines.
The red region is excluded by the LHC direct search and the red lines correspond to 
$\sigma_\text{prod} (p p \to H/A) \times \text{Br}(H/A\to\tau\tau)
= 10^{-1},\ 10^{-3}$ [pb] from left to right.
The yellow region is below the neutrino floor of the DM direct detection.
The spin-independent direct-detection cross sections are $\sigma_{SI}=1.0\times 10^{-11}$ [pb] and $2.5 \times 10^{-11} $ [pb] on the yellow lines. The black solid (dashed) lines show $\kappa_{b,\tau} = 1.10, 1.05, 1.01$ from left to right.   
} 
\label{fig-omega}
\end{figure}

Figure~\ref{fig-omega} shows the observables 
in the case that the thermal relic density of the higgsino saturates the observed value: 
$\Omega_\chi h^2 = 0.1188\pm0.001$. In this case, the higgsino mass is about 1.1 TeV. 
In Fig.~\ref{fig-omega}, the sign of the $\mu$-parameter is the same (opposite) 
as the gaugino masses in the right (left) panel. 
The size of the modulus mediation $M_0$ is chosen 
to realize $m_h = 125.09 \pm 0.01$~GeV 
and is shown by the background colors. 
The gray dashed lines correspond to $\Delta_{M_0} = 400, 300, 250$ 
from bottom to top,   
where $M_0$ is estimated as $M_0 \simeq 6.0, 7.4, 8.6$ TeV, respectively.  
$\Delta_{M_0} \le 300$ is satisfied above the line where $\tan\beta \gtrsim 32\ (38)$ 
when the sign of the $\mu$-parameter is opposite (same) to the gaugino masses. 
The degree of the tuning of $M_0$ is about $0.33\%$ in this region. 
Note that $\mu \simeq 1.1$ TeV means that 
the $\mu$-parameter should be tuned about $0.3\%$-level 
and the required tunings are comparable. 

The red region is excluded by the LHC direct search 
for the extra neutral Higgs bosons decaying to a pair of tau leptons. 
The red lines correspond to 
$\sigma_\text{prod} (p p \to H/A) \times \text{Br}(H/A\to\tau\tau)
= 10^{-1},\ 10^{-3}$ pb from left to right. 
The exclusion limit for $m_A$ is tighter for larger $\tan\beta$ 
because of the larger production cross section 
and branching fraction to a pair of tau leptons.
The current limit is $m_A \gtrsim 1.4$ TeV for $\tan\beta = 50$ 
while there is no limits on $m_A$ for $\tan\beta \lesssim 15$. 
Note that the limits are not so changed, 
even if the higgsino mass is different. 

$\text{Br}(b \to s \gamma)$ and $\text{Br}(B_s \to \mu^+\mu^-)$ 
are deviated from the central values at $2\sigma$-level, 
in the brown and green regions, respectively. 
The uncertainties are calculated 
by combining the uncertainties of the SM prediction and the experimental results.  
The brown and green lines correspond to 1$\sigma$ deviation from the SM prediction.  
The measurement of $\text{Br}(b \to s \gamma)$ 
excludes the light charged Higgs boson region even $\tan\beta$ is small. 
The measurement of $\text{Br}(B_s \to \mu^+\mu^-)$ 
excludes the light charged Higgs boson and heavy higgsino region 
as far as $\tan\beta$ is large, 
and the deviation reaches the 1$\sigma$-level from the current central value 
even if
the extra Higgs bosons are so heavy that 
$\sigma_{H/A}^{\rm prod} \times \br{H/A}{\tau\tau} < 10^{-3}$~pb is predicted.

The black solid (dashed) lines show $\kappa_{b,\tau} = 1.10, 1.05, 1.01$ 
from left to right.  
Note that $\kappa_b \simeq \kappa_\tau$ is predicted
since the deviations are mostly determined by the mixing between the Higgs bosons.  
The deviations from the SM predictions are more than $1\%$ when
the extra Higgs bosons are lighter than 1.5 TeV. 

The yellow region shows that the spin-independent cross section 
is below the neutrino floor. That means that
the yellow region is very difficult to be probed by the direct detections 
even if the dark matter is saturated by the higgsino. 
The yellow lines show the spin-independent cross section 
in the unit of $\times 10^{-11} $ pb.  
We see that the contribution from the heavy Higgs boson exchange 
reduces (enhances) the cross section 
when $\text{sgn} (\mu M_{1,2}) = -1\ (+1)$. 
%
%
%
\begin{figure}[ph!]
 \centering 
 \includegraphics[width=0.48\linewidth]{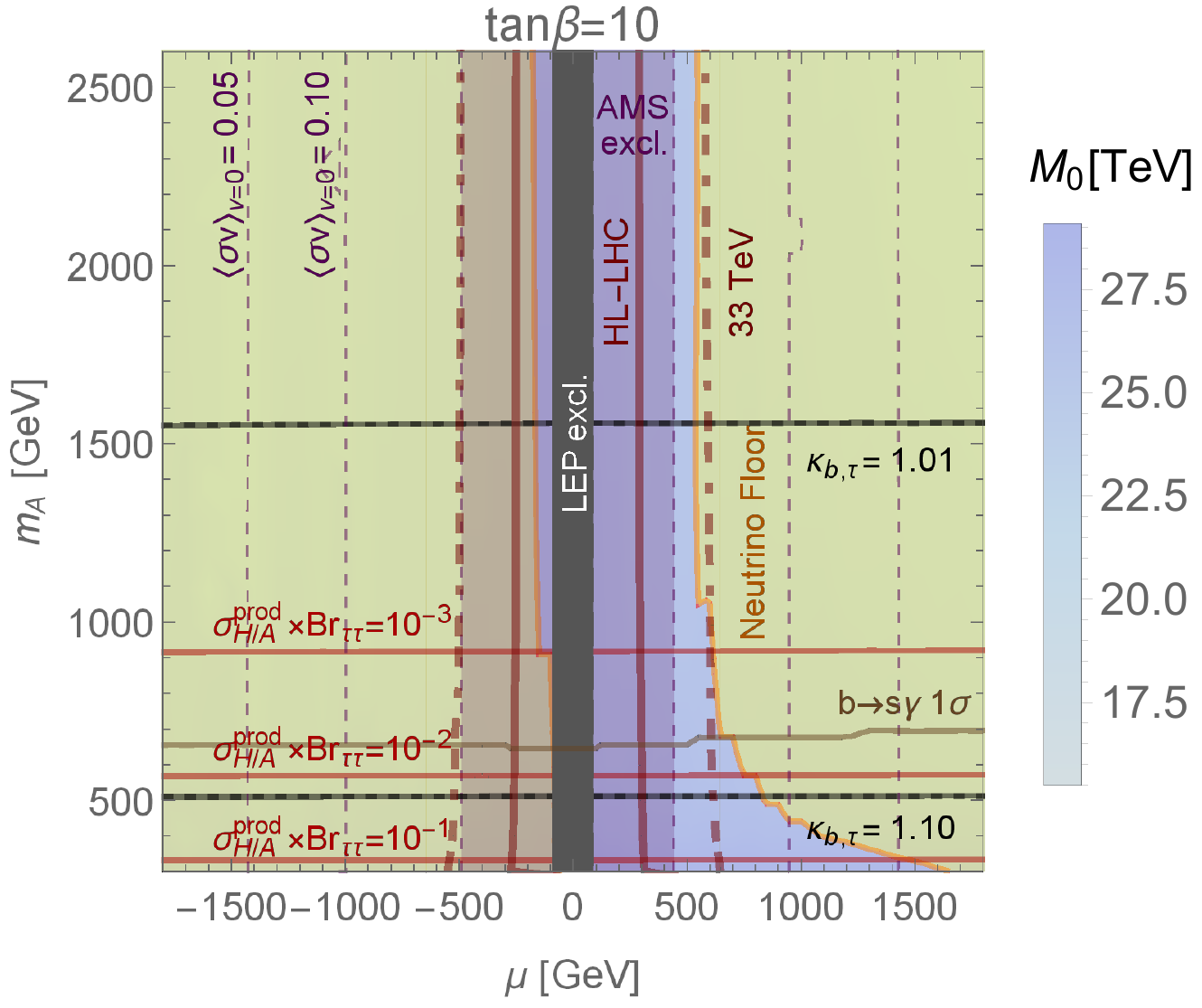}
 \hspace{0.15cm}
 \includegraphics[width=0.48\linewidth]{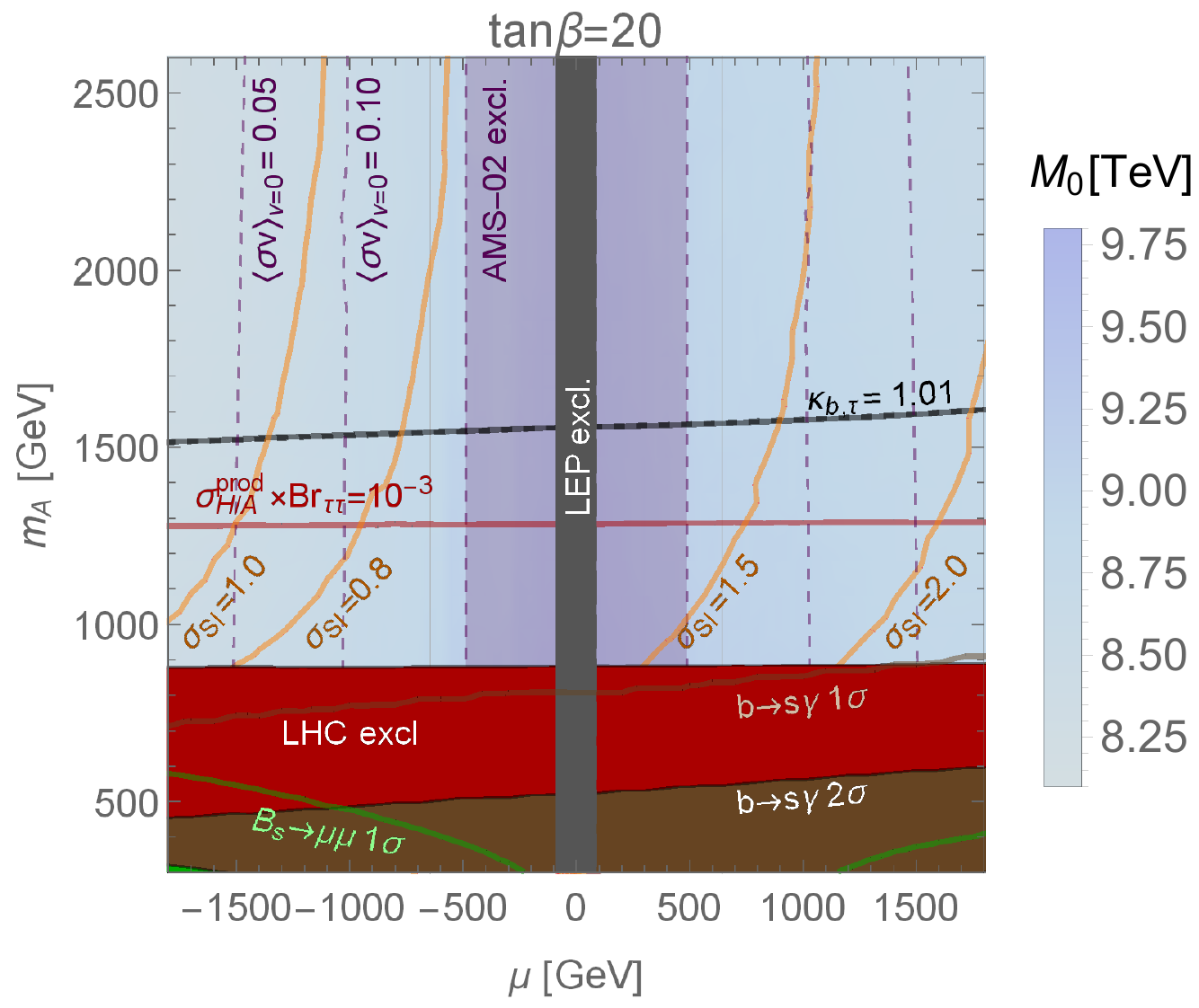}
 \hspace{0.15cm}
 \includegraphics[width=0.48\linewidth]{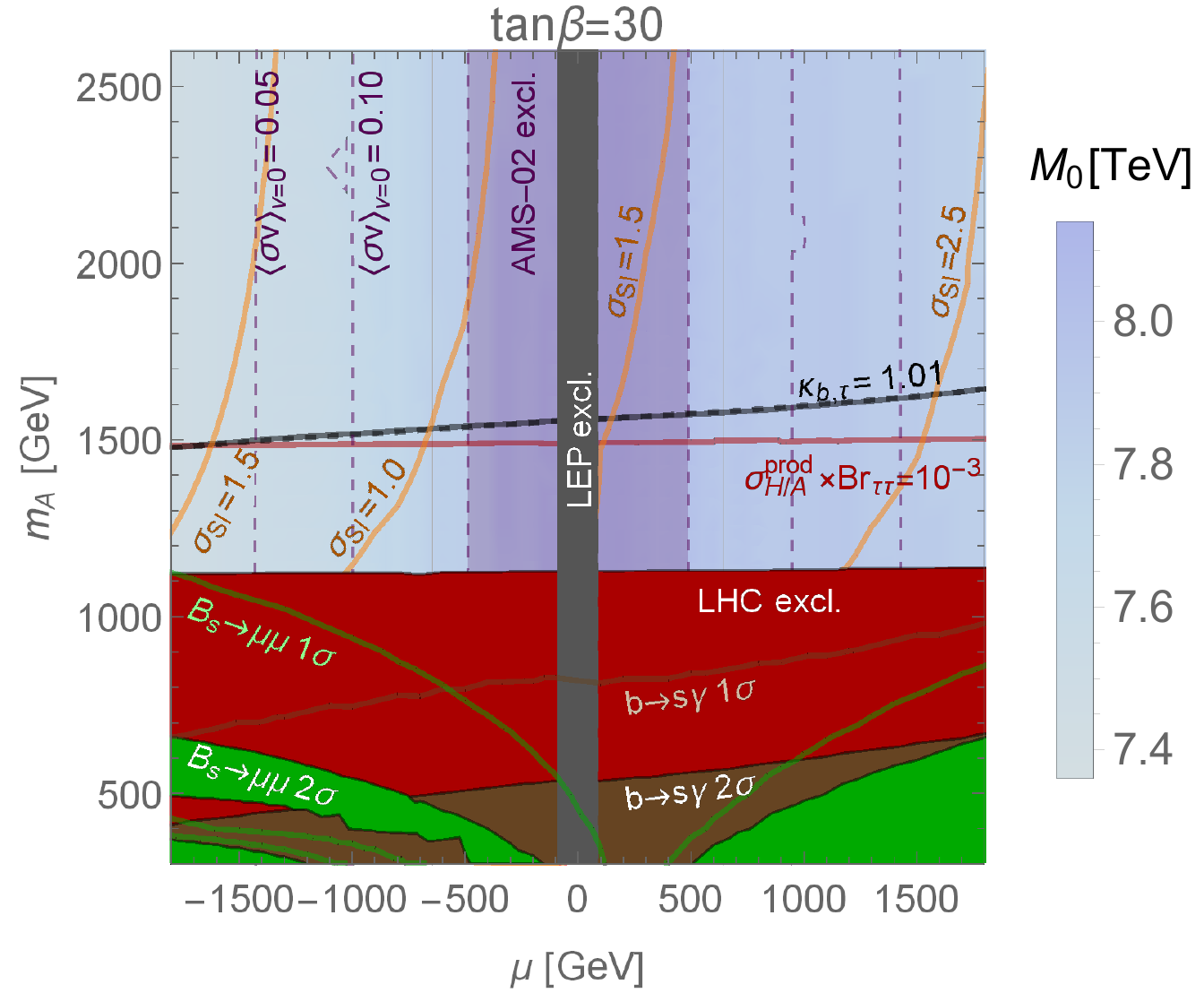}
 \hspace{0.15cm}
 \includegraphics[width=0.48\linewidth]{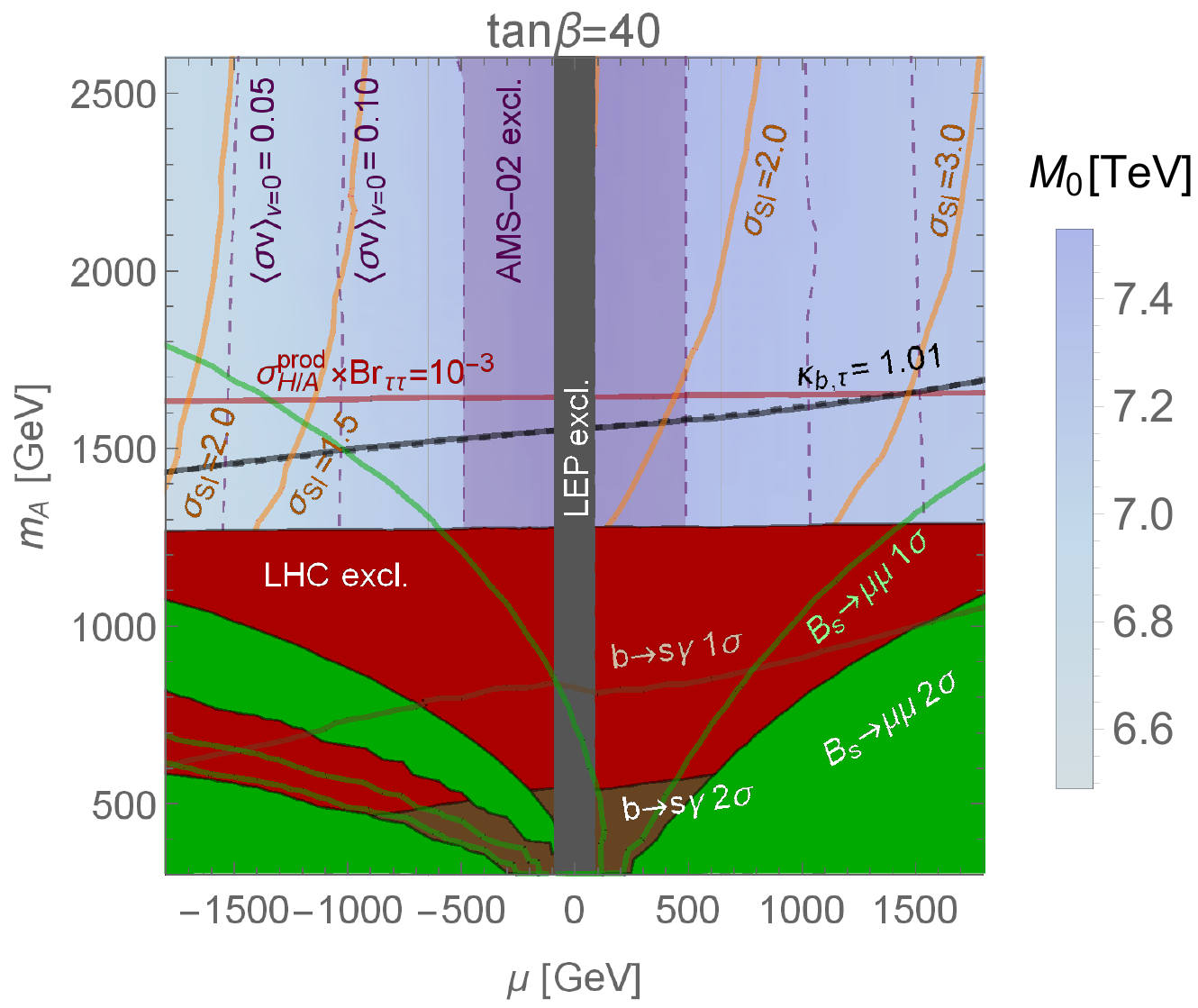}
 \hspace{0.15cm}
 \includegraphics[width=0.48\linewidth]{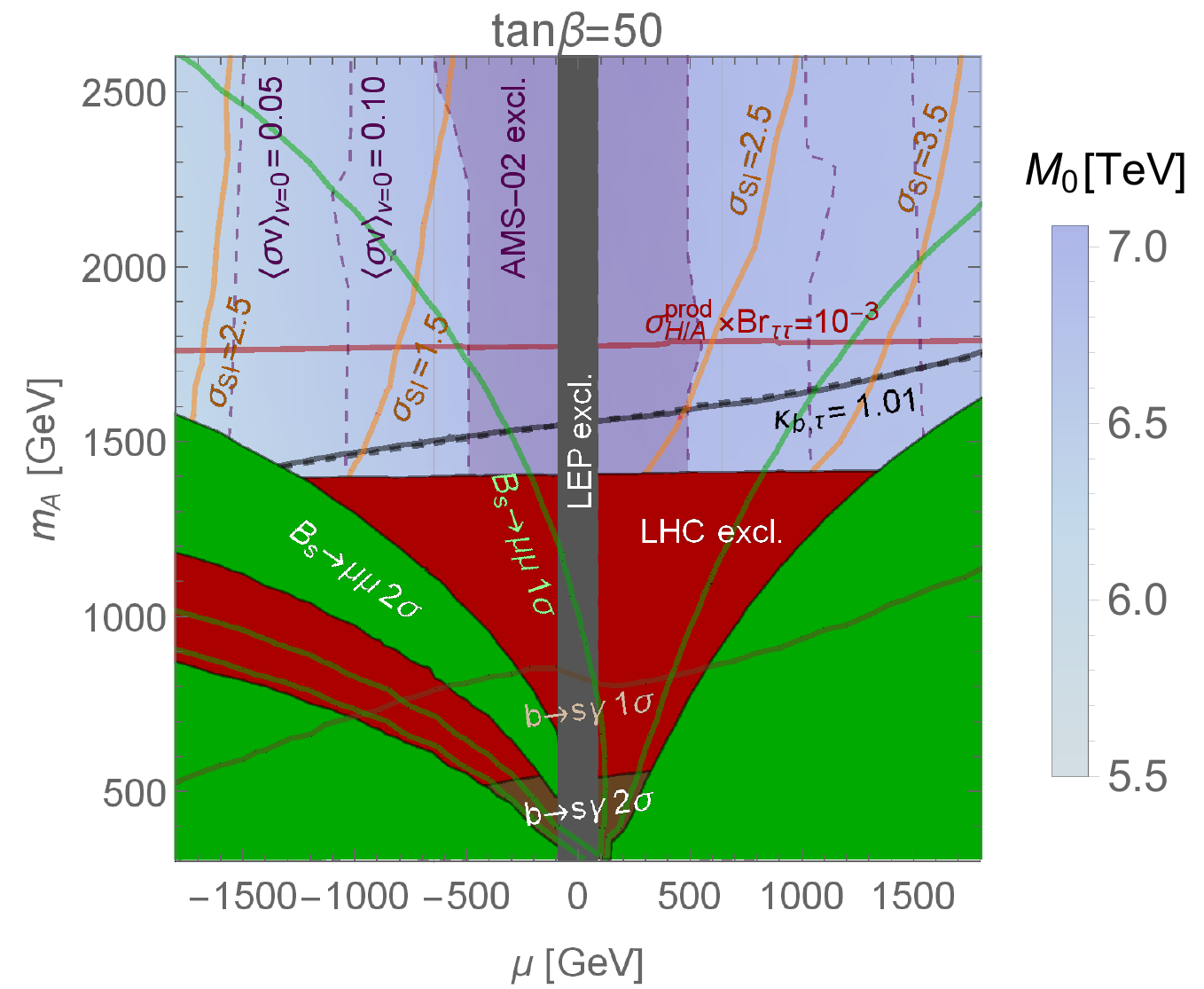}
 \caption{
Values of the observables in the case $\tan\beta=10, 20, 30, 40, 50$.
The meaning of the each line is the same as in Fig.~\ref{fig-omega}.
The gray and purple regions are excluded by the LEP experiment and the AMS-02 experiment. 
The purple dashed lines show the annihilation cross section 
in the unit of $\times 10^{-25} \text{cm}^3/\text{s}$.  
} 
\label{fig-fixedtb}
\end{figure}

Figure~\ref{fig-fixedtb} shows the results on $\mu$-$m_A$ planes with the fixed $\tan\beta$. 
The meanings of the red, brown, green and yellow regions and lines 
are the same as in Fig.\ref{fig-omega}. 
The gray region is excluded by the chargino search at the LEP experiment. 
The purple region is excluded by the AMS-02 experiment 
if the higgsino saturates the dark matter relic density. 
The purple dashed lines show the annihilation cross section in the zero velocity limit 
in the unit of $\times 10^{-25} \text{cm}^3/\text{s}$.  

%
%

The expected upper bounds on $\mu$ from the long-lived higgsino search at
the HL-LHC (the 33-TeV hadron collider) are
described by thick (dashed) red lines in the panel with $\tan\beta = 10$. 
Most of the region with $\Delta_\mu \le 100$ 
could be covered by the long-lived particle search at the 33 TeV hadron collider. 
This region can be covered by the indirect detections for the dark matter 
only if the higgsino dominates the dark matter density. 
Thus the long-lived search is crucial to test the region where $\Delta_\mu \le 100$ 
in the TeV-scale mirage mediation. 
Unfortunately, the long-lived particle search could not probe our scenario 
for larger $\tan\beta$ because of the lighter gaugino masses. 

When $M_0 \lesssim 10$ TeV is satisfied, 
$m_A, \mu$ are expected to be below sub-TeV 
due to the vanishing modulus mediation for the Higgs soft masses. 
The current limit on the parameter space comes 
from the direct search for the extra Higgs bosons
except large $\mu$-parameter region with $\tan\beta = 50$, 
where the bound from $\brbsm$ becomes severe. 
Those limits are highly dependent on $\tan\beta$. 
Note that the 125-GeV Higgs boson couplings 
to the bottom quarks and the tau leptons 
deviate from the SM value at $1\%$-level when $m_A \lesssim 1.5$ TeV. 

The yellow lines show the spin-independent cross section. 
The cross section is under the neutrino floor at $\tan\beta = 10$, 
while whole region in the figures will be covered by the future experiment 
if the higgsino is the dark matter.  
The cross section depends on the heavier Higgs boson mass significantly 
and it increases (decreases) as $m_A$ increases, 
when $\mu$ is same (opposite) sign as the gaugino masses 
as can be read from Eq.~(\ref{eq-sigSI}). 

The annihilation cross section at zero-velocity limit 
is above $1.0\times10^{-26} {\rm cm}^3/{\rm s}$ at $|\mu|\lesssim1.0$ TeV,  
so that the future indirect detections could cover the region 
where the degree of tuning the $\mu$-parameter is $\mathcal{O}(0.1)$\%-level. 
The event rate of the annihilation reduces significantly 
as the abundance of the higgsino decreases  
because the event rate is suppressed 
by a square of the rescaled factor $\xi = \Omega_\chi / \Omega_\text{DM}$.

\section{Conclusion}
In this paper, we have studied the TeV-scale mirage mediation scenario 
that can explain the EW scale naturally without conflict with the current experimental results. 

A specific feature of the mirage mediation is that 
the RG effects for the soft parameters given by the modulus mediation are compensated 
by the anomaly mediated contributions 
below the unification scale, and both contributions cancel out at the {\it mirage scale}. 
This cancellation also happens in the soft SUSY scalar masses as well as the gaugino masses, 
if the modulus mediation respects the condition~Eq.~(\ref{eq-mircond}), 
namely the {\it mirage condition}.  
Exploiting this unification feature, 
$m^2_{H_u}$ and the $\mu$-parameter, 
which are relevant to the EW symmetry breaking, 
can be smaller than the other soft parameters when the modulus mediated contributions 
to the Higgs soft parameters are vanishing at the gauge-coupling unification scale ($M_U$)~\footnote{ We define $c_{H_{u,d}}$ at $M_U$ and require $c_{H_{u,d}}=0$ in Eq. (\ref{eq-choiceC}). }
and the {\it mirage scale} is near the TeV scale. 

The {\it mirage condition} leads small $m^2_{H_u}$, 
but it also predicts small top squark mixing. 
Then the top squark mass should be heavier than about 5 TeV to realize 
the 125-GeV Higgs. 
This fact indicates that all superpartners except the higgsino are 
also heavier than 5 TeV in our scenario
and consistent with the stringent bounds 
on masses of superpartners with color at the LHC. 

The assignment Eq.~(\ref{eq-choiceC}) leads light extra Higgs bosons, 
and then the effective theory below the SUSY scale 
is like the type-II two-Higgs double model (2HDM) accompanied with the higgsinos. 
The apparent difference from the 2HDM is the existence of the higgsinos. 
The higgsino could be the dark matter and 
can be detected by the dark matter experiments. 
The higgsino in our scenario will be searched by 
not only the lepton collider but also the hadron collider exploiting disappearing tracks 
if $\tan\beta \lesssim 10$.  
The light higgsino also influences physical observables such as $\brbsm$, $\brbsg$ 
and the decay of the 125-GeV Higgs boson, 
so that these would be different from the predictions in the 2HDM.
The differences are getting significant as $\tan\beta$ increases.

We emphasize that the TeV-scale mirage mediation discussed in this paper 
still holds motivations of the low-scale supersymmetry. 
The EW scale can be explained by tuning the parameters at $0.3\%$-level. 
The {\it mirage unification} features allows us to control the soft parameters at low energy 
and the parameters relevant to the EW symmetry breaking 
can remain below the TeV-scale. 
In our scenario, the higgsino is a candidate for the dark matter.  
The extra Higgs bosons and the higgsinos are expected to be below TeV-scale, 
so that they can be explored by the current and upcoming experiments in 
both the direct and indirect ways. 


\afterpage{\clearpage}
\subsection*{Acknowledgments}
The work of J. K. is supported by Grant-in-Aid for
Research Fellow of Japan Society for the Promotion of
Science No. 16J04215. 
The work of Y. O. is supported by Grant-in-Aid for Scientific research 
from the Ministry of Education, Science, Sports, and Culture (MEXT), Japan, 
No. 17H05404. 

\appendix

\end{document}